\documentclass[aps,prl,twocolumn,showpacs,amssymb,notitlepage]{revtex4-1}

\usepackage{color}
\usepackage{graphicx}
\usepackage{amsmath}
\usepackage{amsfonts}
\usepackage{amscd}
\usepackage{epsfig}
\usepackage{amssymb}
\usepackage{tabularx}
\usepackage{calligra}

\renewcommand{\a}{\alpha}
\renewcommand{\b}{\beta}
\newcommand{\g}{\gamma}
\newcommand{\e}{\epsilon}
\newcommand{\m}{\mu}

\renewcommand{\d}{\delta}

\newcommand{\w}{\omega}

\newcommand{\p}{\partial}

\renewcommand{\L}{{\cal{L}}}

\newcommand{\M}{{\cal M}}

\begin{document}

\title{Non-modal linear stability of the Schwarzschild black hole}
\author{Gustavo Dotti} \email{gdotti@famaf.unc.edu.ar}
\affiliation{Facultad de Matem\'atica,
Astronom\'{\i}a y F\'{\i}sica, Universidad Nacional de C\'ordoba,\\
Instituto de F\'{\i}sica Enrique Gaviola, Conicet.\\
Ciudad Universitaria, (5000) C\'ordoba, Argentina}
\email{gdotti@famaf.unc.edu.ar}

\begin{abstract}
A proof is given that the space $\L$ of  solutions of the linearized  vacuum  Einstein  equation around a Schwarzschild
black hole is parametrized by two scalar fields,  which are   gauge invariant combinations of  perturbed 
algebraic and differential invariants 
of the Weyl tensor and encode  the information on the odd ($-$) and even ($+$) sectors $\L_{\pm}$.
These fields  measure the distortion of the geometry caused by a generic perturbation and 
 are shown to be pointwise bounded 
on the outer region $r \geq 2M$.
\end{abstract}

\maketitle

\section{Introduction}
The formation of black holes in gravitational collapse is a fascinating prediction of Einstein's   General Relativity, 
which  is backed by a growing astrophysical evidence of their existence and abundance in the Universe. 
The mathematical modeling of these objects, however, is a field where  a number of crucial problems are still open,  
the most salient ones being ruling out the alternative of formation of a naked singularity in gravitational
collapse, that is, establishing the validity of some form of Penrose's cosmic censorship conjecture, and 
proving the stability and, thus, the physical relevance,  of the known black hole solutions. 
Although the first  such a  solution  was found  by Karl  Schwarzschild a few months 
after Einstein's field equations were published, its stability under perturbations has not yet been fully established.  
Even the proof of the  linear stability of the Schwarzschild black hole remains incomplete, 
as only its 
  modal linear stability, 
defined  as  the boundedness of the potentials that generate isolated harmonic modes, has been proved.
In this letter we show that  the geometric information 
of the most general linear perturbation is encoded in two spacetime scalar fields $\Phi_{\pm}$  made from perturbed curvature  invariants.
These fields  measure the distortion of the geometry and are shown 
 to be   pointwise bounded. 
Besides giving  a definite
answer to the problem of linear stability of the Schwarzschild black hole, the
techniques we introduce are likely to apply to the rotating Kerr black hole \cite{ba}.
\\
The  linearized Einstein's equation (LEE) 
describes the propagation of a perturbation $\d g_{\a \b}$ of the metric as a wave  on the background spacetime $(M,g_{\a \b})$
\begin{equation} \label{lee}
\nabla^{\g} \nabla_{\g} \d g_{\a \b} + \nabla_{\a} \nabla_{\b} (g^{\g \d} \d g _{\g \d}) - 2 \nabla^{\g} \nabla_{(\a} \d g_{\b) \g} =0.
\end{equation}
Trivial solutions of this equations are obtained by 
relabeling the points of $M$ by means of   an  infinitesimal diffeomorphism 
$V^{\a}$, under which 
\begin{equation} \label{gt}
g_{\a \b} \to g_{\a \b} + \pounds_V   g_{\a \b} = g_{\a \b} + \nabla_{\alpha} V_{\beta} +
\nabla_{\beta} V_{\alpha}. 
\end{equation}
Two solutions of (\ref{lee}) that differ by 
$ \pounds_V   g_{\a \b} = \nabla_{\a} V_{\b}+\nabla_{\beta} V_{\alpha}$ are, therefore, physically equivalent; 
 this is the gauge invariance of linearized gravity. The problem of linear stability of a stationary vacuum metric $g_{\a \b}$ 
is that of finding out whether the effects on the geometry of a solution of (\ref{lee}) are bounded 
by the initial data of the wave or not.\\
The study of the linear stability of a Schwarzschild black hole $\M$ has a long history that dates
 back to the 1957 pioneering work 
of Regge and Wheeler \cite{Regge:1957td}, where the spherical symmetry of the background was used to 
split  the metric perturbation  into what were termed even  ($+$) and odd  ($-$)   modes of 
 harmonic numbers $(\ell,m)$. Since  these modes do not mix at the linear level, pure modes are  analyzed.
A master variable $\phi_{(\ell,m)}^{\pm}$ defined on the $(t,r)$  space (the orbit space 
$\M / SO(3)$),  
is  extracted   for single modes and the LEE is  reduced to a two dimensional scalar  wave equation on the orbit space:
\begin{equation} \label{rwe}
\frac{\partial^2 \phi_{(\ell,m)}^{\pm}}{\partial t^2}+  {\cal H}_{\ell}^{\pm} \phi_{(\ell,m)}^{\pm}=0, \; \;
 {\cal H}_{\ell}^{\pm}  \equiv  - \frac{\partial^2 }{\partial x^2} +
f \;  U_{\ell}^{\pm}
\end{equation}
where   $f=1-2M/r$ and 
$x$ is a ``tortoise" radial coordinate, defined by $dx/dr= 1/f$.
For the odd modes equation  (\ref{rwe}), known as the Regge-Wheeler equation,  was worked out in   \cite{Regge:1957td}, 
and  the potential is
\begin{equation} \label{potodd}
 U_{\ell}^{-}= \left(\tfrac{\ell (\ell+1)}{r^2} -\tfrac{6M}{r^3}\right).
\end{equation}
The more intricate system of even perturbations was simplified to the form (\ref{rwe})
by Zerilli in 1970 \cite{Zerilli:1970se}, the potential for these modes is 
\begin{equation} \label{poteven}
 U_{\ell}^{+}= \frac{\mu^4 r^2[(\mu^2+2)r+6M]+36M^2 (\mu^2 r+2M)}{r^3 (\mu^2 r+6M)^2}, 
\end{equation}
$\mu^2 = (\ell+2)(\ell-1)$. 
 Note that  $V_{\ell}^{\pm} \equiv f U_{\ell}^{\pm}$ are positive and go to zero 
  at both the 
$x \to -\infty$ (black hole horizon) and $x \to \infty$ (spacelike infinity) limits, thus ${\cal H}_{\ell}^{\pm}$ in (\ref{rwe})  is
formally equivalent to a  positive definite quantum Hamiltonian acting on the space of square integrable functions of $x$.
The gauge invariance of  $ \phi_{(\ell,m)}^{\pm}$ was proved by Moncrief in  \cite{Moncrief:1974am};  
Sengupta and Gerlach   \cite{Gerlach:1979rw} showed how to  recast the Regge-Wheeler equations
and their relation to odd  metric perturbations   as covariant equations in the two dimensional orbit space
(for a quick review, see Section II of \cite{Sarbach:2001qq}, where the even sector is 
worked out).
 If we write the Schwarzschild metric as \cite{Gerlach:1979rw}
\begin{equation} \label{ssm}
g_{\a \b} dz^{\a} dz^{\b} = g_{ab} dx^a dx^b + r^2(x) \hat g_{ij} dy^i dy^j,
\end{equation}
where $ \hat g_{ij} dy^i dy^j = d \theta^2 + \sin^2 \theta \; d\phi^2$ is the metric on the unit sphere
and  $g_{ab} dx^a dx^b$ the Lorentzian metric on the orbit space (given by $-(1-2M/r) dt^2 + dr^2 /(1-2M/r)$ in 
Schwarzschild coordinates), and let $g^{ab}, D_b, \epsilon_{ab}$ be the inverse metric, covariant derivative 
and volume form of the orbit space and  $\hat g{}^{ij}, \hat D_k, \hat \epsilon_{ij}$ those of the unit sphere, then 
the Regge-Wheeler and Zerilli equations (\ref{rwe}) read
\begin{equation} \label{1ssh}
g^{a b}D_a D_b   \phi_{(\ell,m)}^{\pm} -  U_{\ell}^{\pm} \phi_{(\ell,m)}^{\pm} = 0.
\end{equation}
To reconstruct the metric perturbation,  a {\em real} orthonormal basis of  spherical harmonics  $S_{(\ell,m)}$ is used:
\begin{equation} \label{ssh}
\hat D^i \hat D_i S_{(\ell,m)} = -\ell (\ell+1) S_{(\ell,m)}.
\end{equation}
Odd $\ell>1$ modes  in the Regge-Wheeler gauge are 
\begin{align} \label{oddmodes}
\delta g_{ai}^{(\ell,m,-)} &=  [\epsilon_{ac} g^{cb} D_b (r \phi_{(\ell,m)}^-)]  [\hat \epsilon_{ki} \hat g^{kj} \hat D_j S_{(\ell,m)}],\\  \label{oddmodes2}
\delta g_{ab}^{(\ell,m,-)} &=0, \;\;   \delta g_{ij}^{(\ell,m,-)}=0
\end{align}
where $\phi_{(\ell,m)}^-$ is an arbitrary real solution of (\ref{1ssh}).
For $\ell=1$ 
\begin{equation}
J_{m}{}^l = \hat g^{li}  [\hat \epsilon_{ki} \hat g^{kj} \hat D_j S_{(\ell,m)}]
\end{equation}
 is a basis of  Killing vectors on the sphere, 
 tangent to rotations around three orthogonal axes
 (e.g., $J_3= \p/ \p \phi$), and a gauge  
can be found such that the only non zero components of 
an arbitrary linear combination of the  $\delta g_{ai}^{(1,m,-)}$ 
  in standard coordinates are
  (see equation (11) in \cite{Sarbach:2001qq})
\begin{equation} \label{l1odd}
\delta g_{t i}^{(1,-)} = -r^{-1} \textstyle{\sum_{m=1}^3} \; \delta a^{m} J_{m}{}_i.
\end{equation}
This perturbation corresponds to turning on an infinitesimal angular momentum in the direction 
$\delta \vec{a}=(\delta a^1, \delta a^2, \delta a^3)$.
 The odd $\ell=0$ mode is void, whereas the even $\ell=0$ mode amounts to a change of the black hole mass
\begin{equation}  \label{l0even}
\delta g_{t t}^{(0,+)}  =  \tfrac{2 \delta M}{r}, \;\; 
  \delta g_{rr}^{(0,+)} =  \tfrac{2 \delta M}{ r (1-2M/r)^{2}}
\end{equation}
Even $\ell=1$ modes are pure gauge, whereas higher even modes  are involved expressions on 
$ (\phi^{+}_{(\ell,m)}, S_{(\ell,m)})$ that we need not spell out here  (see, e.g.,\cite{Gerlach:1979rw,Sarbach:2001qq}).\\
The Zerilli and  Regge-Wheeler equations
allow us to analyze isolated modes  and to establish a basic notion of linear stability,  based on the
 fact that   they  
admit separable solutions of the form $\phi = e^{\a t} \chi(x)$ (we suppress indices for simplicity).
 From equation (\ref{rwe}) it  can easily be shown that 
$\a$ must be purely imaginary, $\a =i \w$, since $\a^2 \chi = - {\cal H} \chi$ and 
$ {\cal H} $ is positive definite. Since the perturbation of  any geometric field is 
obtained by applying a linear differential operator
to $\delta g_{\a \b}$,  it will also be oscillatory and, thus, bounded in time.
Exponential growth for non separable solutions of (\ref{rwe}) can be ruled out using the positive definite conserved energy of 
(\ref{rwe}). An integral bound of the form 
$\int | \phi_{(\ell,m)}(t,x) |^2 dx  \leq C_{(\ell,m)}$
 was obtained in
\cite{wald1056}, where the possibility of unbounded growth in narrowing intervals 
was also ruled out by  proving
 that a pointwise bound can be placed on the $\phi^{\pm}_{(\ell,m)}$ 
\begin{equation} \label{wald}
| \phi^{\pm}_{(\ell,m)} |  \leq K_{(\ell,m)}^{\pm},  \;\; r>2M,
\end{equation}
with $K_{(\ell,m)}^{\pm}$  a constant obtained from the $(\ell,m,\pm)$ piece of the initial data.

\section{Non-modal stability}
 The most general   linear perturbation of the Schwarzschild black hole is of the form 
\begin{multline} \label{series}
 \delta g_{\a \b}^{(1,-)}(\delta \vec{a}) + \delta g_{\a \b}^{(0,+)}(\delta M) +\sum_{m, P=\pm \atop \ell \geq 2}  \delta g_{\a \b}^{(\ell,m,P)}[\phi,S]
\end{multline}
 The first two terms in (\ref{series}), given in  (\ref{l1odd})-(\ref{l0even}), are deviations within the Kerr family and 
 are time independent: no dynamical process can lead to a  change in mass or angular 
momentum  in the linear regime, these processes show up at second order \cite{Gleiser:1997fk}. The series  
 in (\ref{series}), whose 
completeness  follows from 
the theorems in section 2.3 of \cite{Ishibashi:2011ws},  
does not contribute to any of these charges and encodes 
 the dynamics of the perturbation.
The $ \phi^{\pm}_{(\ell,m)}$ are  an infinite set of  potentials whose
derivatives enter individual terms in this series, 
and two extra derivatives must be taken to calculate the perturbed Riemann tensor and analyze the 
effects of the perturbation. \\ In order to evaluate whether a perturbation ``grows big'' or not, 
we  first need  to parametrize the space $\L$ of solutions of the LEE
 (\ref{lee})  with 
geometrically meaningful quantities.
For this purpose, it is important  to understand how the different modes in (\ref{series}) behave
under the action of the symmetry group of the Schwarzschild metric. The
 isometries of the background 
 commute with the 
LEE, equation (\ref{lee}). When 
applied to a linearized solution $g_{\a \b} + \d g_{\a \b}$,  they keep $g_{\a \b}$  fixed while  acting on $\d g_{\a \b}$
as linear operators  in  $\L$.  The isometry group of  the Schwarzschild metric 
is $
{\mathbb R}_t  \times T   \times SO(3) \times P$, where ${\mathbb R}_t$ is the subgroup 
of time translations,  $SO(3)$ are the proper rotations, and  $T$ and $P$ are the ${\mathbb Z}_2$ subgroups of time inversion, 
   $T: (t,r,\theta,\phi) \to (-t,r,\theta,\phi)$, and parity transformation 
$P: (\theta,\phi) \to (\pi-\theta,\phi+\pi)$. 
The $(\ell,m)$ labels  
 are attached to modes constructed from the $S_{(\ell,m)}$ spherical harmonics and their first and second derivatives. These derivatives 
are the components of 
eigentensors of the Laplacian on the sphere  \cite{Ishibashi:2011ws}. 

The ``square angular momentum" operator 
\begin{equation} \label{sam}
{\mathbf J}^2 = \textstyle{\sum_{m=1}^3}  \; ( \pounds_{J_m})^2
\end{equation}
acting on (\ref{oddmodes}) gives 
\begin{equation} \label{Jg}
{\mathbf J}^2 \delta g_{\a \b}^{(\ell,m,\pm)} = -\ell (\ell +1) \delta g_{\a \b}^{(\ell,m,\pm)}, 
\end{equation}
and if  we used the standard spherical harmonics $Y_{(\ell,m)}$  instead of a real basis,  
$ \delta g_{\a \b}^{(\ell,m,\pm)}$
would also be a (complex) eigentensor of  $\pounds_{J_3}$, with eigenvalue $im$. 
The meaning of ``even" and ``odd" modes (introduced with quotes in the original work \cite{Regge:1957td}) 
is more obscure, it tells us  whether the perturbation behaves as a scalar field or not
\begin{equation}
P_*  \delta g_{\a \b}^{(\ell,m,\pm)} = \pm (-1)^{\ell}
 \delta g_{\a \b}^{(\ell,m,\pm)}.
\end{equation}
A more significant interpretation can be given, as we now 
proceed to explain.
For vacuum spacetimes, every  algebraic invariant of the Riemann tensor can be written 
as a polynomial in four basic invariants:
\begin{align} \label{invs}
Q &=  \tfrac{1}{48}\ ( C_{\a \b \g \d} + i C_{\a \b \g \d} ^* ) C^{\a \b \g \d}  = Q_++  i Q_-\\ \nonumber
C &:= \tfrac{1}{96} ( C_{\a \b}{}^{\g \d} + i C_{\a \b}^* {}^{\g \d} ) C_{\g \d}{}^{\e \m} C_{\e \mu}{}^{\a \b} = C_+ + i C_-, 
 \end{align}
where  $C_{\a \b \g \d}$ is  the Weyl tensor  and 
 $C_{\a \b \g \d} ^* := \frac{1}{2} \epsilon_{\a \b}{}^{\e \m} C_{\e \m \g \d}$ its dual. Note that $\epsilon_{\a \b \g \d} $,  
the volume form of spacetime,  is odd under parity, 
 $P_*  \; \epsilon_{\a \b \g \d}  = -\epsilon_{\a \b \g \d} $ and that $Q_-$ and $C_-$ are 
``pseudoscalars", i.e.,  their construction requires  the volume form 
besides the metric, they are orientation-dependent, and pick up an extra minus sign under $P$, 
their $(\ell,m)$ piece transforming as $(-1)^{\ell+1}$.
For the Schwarzschild spacetime,
\begin{equation} \label{si}
Q = \tfrac{M^2}{r^6}, \;\;\; C =  \tfrac{M^3}{r^9}, 
\end{equation}
the vanishing of the pseudoscalars being forced by the  facts that they must be odd under $P$ but cannot 
depend on the angular variables (since they must vanish under $\pounds_{J_k}$).
In addition to $Q$ and $C$, differential invariants of the Weyl tensor are required to fully characterize
a vacuum metric, the simplest one being
\begin{equation}\label{de}
X = \tfrac{1}{720} (\nabla_{\epsilon} C_{\alpha \beta  \gamma \delta}) ( \nabla^{\epsilon}
 C^{\alpha \beta  \gamma \delta})= \tfrac{M^2}{r^9} (r-2M)\\ \nonumber
\end{equation}

Under a perturbation, the first order variation  $\delta I_{\pm}$ of a (pseudo) scalar  
  invariant  $I_{\pm}$
of the  Weyl tensor (such as $Q, C, X$ and $Y$ above)  is a  linear functional of $\delta g_{\a \b}$ that 
commutes with all symmetries, and then 
\begin{multline}
(-1)^{\ell} \delta I_+ [  \delta g_{\a \b}^{(\ell,m,-)} ] =  P  \delta I_+ [  \delta g_{\a \b}^{(\ell,m,-)} ] \\  
= \delta I_+ [  P_* \delta g_{\a \b}^{(\ell,m,-)} ] = 
  - (-1)^{\ell} \delta I_+ [  \delta g_{\a \b}^{(\ell,m,-)} ],
\end{multline}
which implies that     $\delta I_+ [  \delta g_{\a \b}^{(\ell,m,-)} ] =0$. Similarly,  $\delta I_- [  \delta g_{\a \b}^{(\ell,m,+)} ]$ must be zero.
Thus, (odd) even perturbations can be better characterized as those exciting perturbations of curvature (pseudo) scalars.\\
 If we calculate  $\delta Q_- [  \delta g_{\a \b}^{(\ell,m,-)} ]$ from (\ref{series}), 
 we get a rather complicated expression which
simplifies if we make use of the LEE  together with  their derivatives, leaving a 
 strikingly simple expression:
\begin{equation} \label{dqm}
\delta Q_- = \tfrac{-6M^2 }{r^7}  S_{(1,m)}  \delta a^m + \tfrac{3 M}{r^6} \sum_{\ell>1,m} 
 \tfrac{(\ell+2)!}{(\ell-2)!}  \; \phi^{-}_{(\ell,m)} S_{(\ell,m)}
\end{equation}
The above equation shows that the local curvature pseudoscalar $\delta Q_-$ encodes 
all the information carried by the most general odd perturbation, as the non local quantities $\delta \vec{a}$ and 
 $ \phi^{-}_{(\ell,m)}$ can be recovered by integrating $\delta Q_-$ against  $S_{(\ell,m)}$  on the sphere.
 Moreover,  (\ref{dqm}) together with  (\ref{1ssh})-(\ref{ssh}) implies that
\begin{equation} \label{fim}
\Phi_- \equiv \tfrac{r^5}{3 M} \; \delta Q_-
\end{equation}
satisfies 
a simple {\em four dimensional}  wave equation on the Schwarzschild spacetime:
\begin{equation} \label{oce}
\left[ \nabla_{\a} \nabla^{\a}  +\frac{8M}{r^3} \right] \Phi = 0
\end{equation}
 ($\Phi=\Phi_-$). This equation is also satisfied by the potential  $\Phi_-^{o}  =\sum_{\ell \geq 2, m}  \tfrac{\phi^{-}_{(\ell,m)}}{r} S_{(\ell,m)}$, 
in terms of which the dynamical terms of the metric perturbation (\ref{oddmodes}) can be compactly written in a covariant way as 
\begin{equation} \label{seed}
  \sum \nolimits_{(\ell>1,m)} \delta g_{\a \b}^{(\ell,m,-)} = \tfrac{r^2}{3M}  {}^*C_{\a}{}^{\g \d}{}_{\b} \nabla_{\g} \nabla_{\delta} (r^3 \Phi_-^{o}).
\end{equation}
 \noindent
{\em Theorem: Non-modal linear stability of the Schwarzschild black hole (odd sector):} \\
The space $\L_-$ of solutions of the LEE around a  Schwarzschild black hole mod gauge transformations is 
parametrized by the gauge invariant  pseudo scalar field  $\delta Q_-$. 
For any perturbation with compact support on Cauchy surfaces of the Kruskal extension, 
\begin{equation} \label{boundo}
\left| \delta Q_- \right| < K_-/r^6
\end{equation}
on the exterior wedge $r \geq 2M$, $K_-$ a constant that depends on the perturbation data on a $t-$slice. \\

{\em Proof:} The only thing that remains is to prove the bound (\ref{boundo}). 
 $\Phi_-$ satisfies (\ref{oce}) which, following 
 \cite{Kay:1987ax}, where the  Klein Gordon equation on the Schwarzschild background  is studied,  can be written
 as $0 =( \p_t^2 - \p_x^2 + V_1 - \hat D^k \hat D_k V_2) (r \Phi)$. 
 We find from (\ref{potodd}) that  $V_1 = -(1-2M/r) 6M/r^3$ and 
$V_2 = (1-2M/r)/r^2$  are both bounded for $r \geq 2M$,  therefore the  proof in Appendix A of  \cite{Kay:1987ax} applies to equation (\ref{oce}),
as well as the symmetry argument in the main text,  and implies  that 
$| \Phi_- | < K'_{-} /r$, from where (\ref{boundo}) follows.
\\

Even perturbations are more difficult to deal with for two reasons. (i) 
The dependence of (\ref{poteven}) on $\ell$  indicates that the 
set of Zerilli functions $\phi^{+}_{(\ell,m)}$ is 
not {\em directly} related to the harmonic components of a four dimensional scalar field; and 
 (ii) although the scalar invariant
$Q_+$  is excited by the even modes, the excitations $\d Q_+$ are not gauge invariant because 
$Q_+$ does not vanish in the background and,  under the gauge transformation (\ref{gt}),
\begin{equation} \label{gt2}
\delta Q_+ \to \delta Q_+ + \pounds_V Q_+ =  \delta Q_+ 
+ V^{r} \partial_{r} Q_+
\end{equation}
and similarly for $C_+$. 
 Problem (ii) is 
absent in the odd sector because $Q_-=0=C_-$. 
To tackle  it,  we  could  substitute $\d Q_+$ with
 any gauge invariant combination of perturbed scalars. However,
when computing $\d Q_+$ and $\d C_+$ in the Regge-Wheeler gauge  we find that  $\d Q_+ / \d C_+ =
 \p_r C_+ / \p_r Q_+$, 
and this fact, together with (\ref{gt2}),  implies that all such  gauge invariants  will vanish under a genuine   perturbation, and so are useless. 
Thus, 
 we need to incorporate {\em differential} invariants,  such as $X$ in (\ref{de}),  which do not 
satisfy simple equations. The simplest gauge invariant combination  of the enlarged set of perturbed scalars is 
\begin{equation} \label{fip}
\Phi_+ = (9M-4r) \d Q_+ + 3r^3 \d X.
\end{equation}
We will use it to measure the effect of even perturbations on the geometry. 
To deal with  (i) we use the factorization property \cite{Chandrasekhar:1985kt} 
\begin{align} \label{facto1}
{\cal H}_{\ell}^{\pm} &= {\cal A}^{\pm}_{\ell} {\cal A}^{\mp}_{\ell}  - E_{\ell}{}^2, \; \; \;
{\cal A}^{\pm}_{\ell} = \pm \frac{\p}{\p x} +W_{\ell}\\
W_{\ell} &= E_{\ell} + \tfrac{6M(r-2M)}{r^2 ( (\ell+2)(\ell-1)r+6M)}, \;  E_{\ell} = \tfrac{1}{12M} \tfrac{(\ell+2)!}{(\ell-2)!}
\end{align}
 Thus ${\cal A}_{+}  \phi^{-}_{(\ell,m)}$ solves the even equation (\ref{rwe}) if  $\phi^{-}_{(\ell,m)}$
solves the odd one.
This suggest that we write   even metric perturbations using  odd potentials through $\phi^{+}_{(\ell,m)} ={\cal A}_{+} \phi^{-}_{(\ell,m)}$.
 A lengthy calculation using  the LEE then reduces 
$\Phi_+$ in (\ref{fip}) to
\begin{align} \nonumber
 -\tfrac{2M \; \delta M}{r^5} &+ \sum_{3 \leq j\leq 7}  \tfrac{M^{j-3}}{r^j}  [ M \p_r \Phi_{(1,j)}
+ M^2 \p_t^2 \Phi_{(2,j)}
+   \Phi_{(3,j)} ],  \\   \label{pos}
 \Phi_{(k,j)} &= \sum_{(\ell\geq 2, m)}  P_{(k,j)} \tfrac{\phi^{-}_{(\ell,m)}}{r} S_{(\ell,m)}, \; 1\leq k \leq 3,
\end{align}
where 
$P_{(k,j)}$ are polynomials in $\ell$. \\

\noindent
{\em Theorem: Non-modal linear stability of the Schwarzschild black hole (even sector):} \\
The space $\L_+$ of solutions of the linearized Einstein equations around a  Schwarzschild black hole mod gauge transformations is parametrized  
by the gauge invariant scalar field $\Phi_{+}$ in  (\ref{fip}). 
For any perturbation with compact support on Cauchy surfaces of the Kruskal extension, 
\begin{equation} \label{bounde}
\left| \Phi_+ \right| < K_+ /r^3
\end{equation}
on the exterior wedge $r \geq 2M$, with $K_+$ a constant that depends on the perturbation data on a $t-$slice.\\

{\em Proof:}  A generic even perturbation is parametrized by  the Zerilli potentials entering $\delta g_{\alpha \beta}^{(\ell\geq 2,m,+)}$, and $\delta M$ 
(see (\ref{series})). In terms of these, using the LEE one finds 
\begin{align} \nonumber
\Phi_+ &= -\tfrac{2M \; \delta M}{r^5} + \tfrac{M}{2 r^4} \sum_{\ell \geq 2} 
\tfrac{(\ell+2)!}{(\ell-2)!} [\p_x + Z_{\ell}(x)] \phi^{+}_{(\ell,m)} S_{(\ell,m)}\\
Z_{\ell} &= \tfrac{\mu r(r-3M)-6M^2}{r^2 (\mu r + 6M)}, \;\; \mu = (\ell-1)(\ell+2). \label{ip2}
\end{align}
By expanding $\Phi_+$ in spherical harmonics we get  $\delta M$ and 
$ [\p_x + Z_{\ell}(x)] \phi^{+}_{(\ell,m)}$, 
from where the  $\phi^{+}_{(\ell,m)}$ can be solved.
The parametrization of $\L_+$ then follows.  
The alternative form (\ref{pos}),  obtained by replacing $\phi^{+}_{(\ell,m)} ={\cal A}_{+} \phi^{-}_{(\ell,m)}$
in (\ref{ip2}), 
 is used to prove (\ref{bounde}). Every  $\Phi_{(k,j)}$ in (\ref{pos}) -and then every  $\p_t^2 \Phi_{(k,j)}$- 
satisfies the wave equation (\ref{oce}), 
its absolute  value is then bounded by a constant times $r^{-1}$, and 
its $r-$derivative is bounded by a constant on the $r\geq 2M$ region, as proved by a direct transcription 
of  the results in Section 3.6 of \cite{Dain:2012qw}.
Then, (\ref{bounde}) follows.\\

In conclusion, we have shown the following. 
(i) The information on arbitrary metric perturbations is contained in $\Phi_{\pm}$ given in (\ref{fim}) and (\ref{fip}).  These are   gauge invariant 
curvature scalars that can be measured locally, unlike Regge-Wheeler and Zerilli potentials, which require integrations on the sphere. 
(ii) For generic perturbations, the initial data  places a pointwise bound for $\Phi_{\pm}$  in the outer region. \\

I would like to thank Sergio Dain, Reinaldo Gleiser, Jacek Jezierski and Robert Wald  for pointing out errors in 
the first version of this manuscript.
This work was partially funded from grants PICT-2010-1387, PIP 11220080102479 and Secyt-UNC 05/B498. 
The grtensor package (grtensor.org) was used to calculate perturbed curvature invariants.

\end{document}